\documentstyle[psfig,conf_iap]{article}

\begin{document}

\heading{
Small Scale Structure in the Universe and the 
Distribution of Baryons at High Redshift} 
\par\medskip\noindent

\author{
Michael Rauch
}
\noindent{\it
Astronomy Department 105-24 \\ 
California Institute of Technology\\
Pasadena, CA 91125, USA \\
e-mail: mr@astro.caltech.edu
}

%
%

\begin{abstract}

\medskip

We discuss briefly the relevance of Lyman $\alpha$ forest observations
for measuring cosmological parameters, comparing the properties of high
z Keck QSO spectra  with those of artifical spectra from hydrodynamic
simulations, based on hierarchical cosmologies.  In particular, we
describe a measurement of the baryon content of the universe obtained
by matching the average opacity of the Lyman $\alpha$ forest from
simulations to the observed one from a new dataset observed with the
Keck telescope. For conservative assumptions about the intensity of the
UV background we obtain a lower limit $\Omega_{b}h^{2}$ $>$ 0.017.
Searching for column density gradients in absorption systems common to
adjacent gravitationally lensed quasar images we test for the presence
of sub-kpc clumpiness which could invalidate the results of simulations
with limited resolution.  We find that in those Ly$\alpha$ forest clouds dominating
the mean absorption, such structure, if present, cannot exceed the 4
percent level (over 100 -- 200 pc).
Extending the study of lensed absorption to higher column
densities we begin to sample the velocity field internal to high
redshift galaxies. Remarkable differences in column density (of order
50 \%) and projected velocity (tens of km/s) between lines of sight
separated by only a few hundred parsecs are found, and we may be
observing structure in the early interstellar medium.

\end{abstract}
%
%
\medskip

It is obvious from the wide range of contributions to this meeting that
observations and especially the theory of the intergalactic medium have
made substantial progress during the past few years. Bright QSO spectra
with signal-noise ratios of order 100 and a resolution of 5 km/s (FWHM)
can now be obtained routinely in a few hours with the Keck and soon
with other large telescopes.  Cosmological hydro-simulations have
advanced our understanding of the Ly$\alpha$ forest phenomenon to a
point where quantitative cosmology with the IGM has come within reach
(see also the contributions by Bond, Croft, Dav\'e, Davidsen, Gnedin, Hui,
Meiksin, Miralda-Escud\'e, Muecket, Weinberg, at this conference) and
even the study of galaxy formation in absorption is now becoming
possible (talks by Haehnelt, Hellsten, Murakami, Steinmetz).

\section{The Lyman $\alpha$ Forest - Observing the Main Baryonic Reservoir}

\begin{figure}[htb]
\centerline{
\psfig{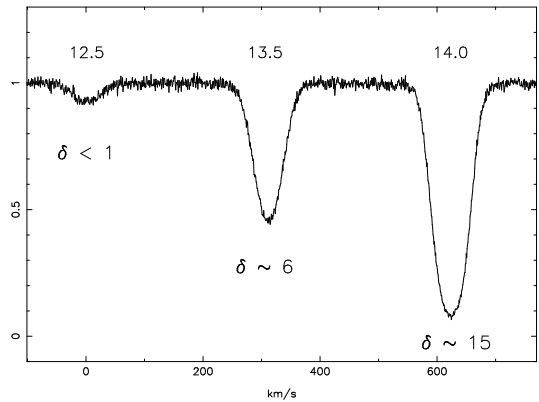}
}
\caption{
\small The observable range of the linear part of the curve of
growth.  The logarithmic column densities and
corresponding typical baryonic overdensities $\delta$ are shown for three artificial 
z=3 HI Ly$\alpha$ lines.
\label{cog}}
\end{figure}

\begin{figure}[htb]
\centerline{
\hskip 2.cm\psfig{file=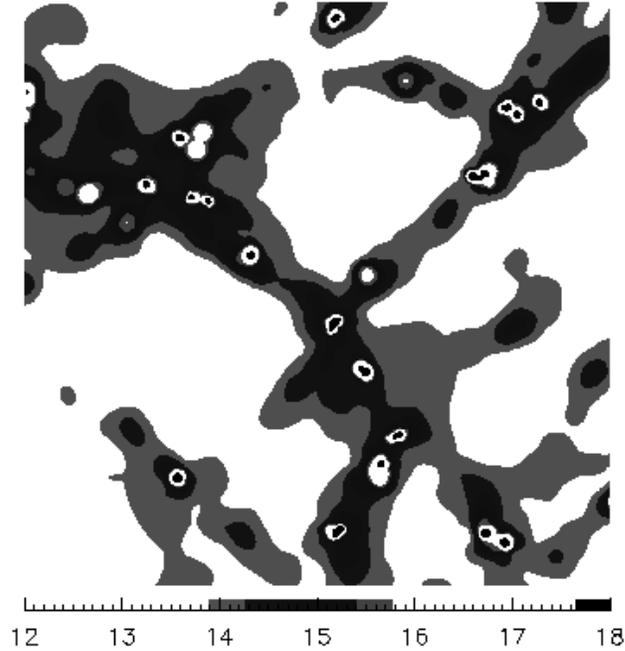,width=14.cm,angle=-0.}
}
\vskip -8.cm
\caption{
\small The projected HI column density distribution in a cube
of size 700 kpc (proper), at z=3.1, from a standard CDM SPH simulation by M. Steinmetz. The contours refer to $\log$N(HI).
\label{densdist}}
\end{figure}

Observations of the Ly$\alpha$ forest provide a very sensitive way for
observing of the cosmic matter distribution and the early stages of
galaxy formation: this is because the {\it linear density regime} of
gravitational collapse near turnaround for galactic scales  (several
hundred kpc at z$\sim 3$) coincides roughly with  the {\it linear part
of the curve of growth} for the Lyman $\alpha$ absorption line, as
shown in fig. \ref{cog}.
The hydro-simulations picture a Ly$\alpha$ forest caused by a
coherent network of filaments, sheets, and knots of gas, in which more
spherical, higher density condensations, galaxies or mini-halos, are
embedded.  The fraction of baryons in Lyman $\alpha$  clouds in CDM
based models is indeed very high \cite{pet}, with (even by
z $\sim$ 2) of order 80\% of all baryons in low column density clouds,
dominated by the range (14 $<$ $\log N$ $<$ 15.5) \cite{me},\cite{hern}.   The spatial arrangement of this baryonic
reservoir is illustrated by fig.  \ref{densdist}.  Column density
contours at a level as high as $10^{14}$ cm$^{-2}$ are stretching
continuously over many hundreds of kpcs.  {\it Most of the baryons are concentrated
in the filaments}, without having collapsed into Lyman limit systems, virialized galaxies or damped Ly$\alpha$ systems yet.

\section{The Baryon Content of the Universe}

The hydrodynamic simulations are illustrative and have
furthered the interpretation of QSO absorption spectra immensely, but
they can also be used for quantitative measurements, e.g., by comparing
large sets of simulated QSO spectra with observed ones.  The problem
is as usual to find observables which can be measured at a reasonable
precision, do not depend sensitively on the shortcomings of the modelling
and at the same time correspond in a unique way to the
ingredients of the simulated model.

One of the most easily measurable quantities is the distribution of
pixel intensities $I$ = $ e^{-\tau}$ (or alternatively, flux decrements
$D = 1 - I$, or optical depths $\tau$), i.e. the amount of light per
unit velocity absorbed by Ly$\alpha$ of intervening HI clouds from the
beam of a QSO.  The optical depth $\tau$ is a measure of the
distribution of the neutral hydrogen in real and velocity space, $\tau
\propto$ d$N_{HI}$/d$v$, where d$N_{HI}$ is the neutral hydrogen column
density spread out over velocity interval d$v$.  The simulations
predict density $\rho$, temperature $T$, peculiar velocity $v_{pec}$,
and thus the ionization state of the gas as a function of the
cosmological model, with the ionizing radiation background, and the
total $\Omega_{b}$ in the universe as free parameters. In the density
range producing most of the absorption the gas is highly ionized, and
photoionization dominates the ionization equilibrium.  Then the optical
depth for absorption is approximately proportional to

\begin{eqnarray} 
\tau \propto \frac{\left(\Omega_b H_0^2\right)^2}
{\Gamma \, H(z)}\, (1+z)^6 \, \alpha(T) \,
 \left(\frac{\rho}{\bar\rho}\right)^2
\left(1+\frac{dv_{pec}}{H(z) dr}\right)^{-1} ~, 
\label{eqn:tau}
\end{eqnarray} 
where $\Gamma$ is the photoionization probability per second,
$\alpha(T)$ the recombination coefficient, $H(z)$ the Hubble constant, $\rho$ the gas density, and 
$dv_{pec}/dr$ the gradient of the peculiar velocity along the line of sight.

Then, for a given temperature, the optical depth scales with  $\Omega_b^2
h^4/(H(z)\Gamma)$.  If we are able to obtain an independent estimate of one of
the parameters e.g., the ionizing flux, we can determine the other,
e.g., $\Omega_b h^{3/2}$ or vice versa.

We have observed a sample of 7 QSOs with the Keck telescope, and
compared the observed distribution of flux decrements $D$ = $1 -
e^{-\tau}$ with the predicted distribution from an Eulerian hydro-simulation of
a CDM+$\Lambda$ universe \cite{cen}, and a standard CDM
universe from an SPH simulation \cite{hern}.  The
measurement consists of applying a suitable global scaling to the
simulated optical depths  such that the mean flux decrements agree
between simulation and observation for each of three redshift bins,
\begin{eqnarray}
\overline{D_{obs}}(z) = \overline{D_{sim}}(z),\ \ \ \ \ \  \ z = 2,3,4.
\end{eqnarray}

\begin{figure}[htp]
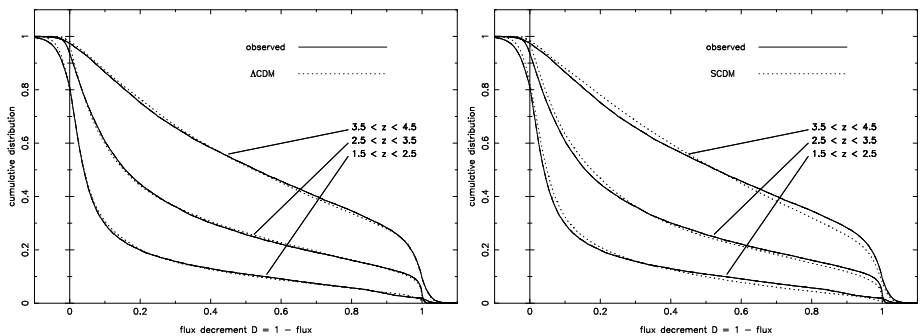

\centerline{
\psfig{file=rauchF3.ps,width=6.cm,angle=-90.}
\psfig{file=rauchF4.ps,width=6.cm,angle=-90.}}
\caption{\small left: Comparison between observed (solid line) and simulated
(best fit: dotted line) cumulative flux decrement distribution for the
CDM+$\Lambda$ model of \cite{cen}. right:  Same statistic for
the standard CDM SPH simulation of \cite{hern}) \label{datsim}}
\end{figure}

Observed and simulated cumulative flux decrement distributions are
shown in fig. \ref{datsim}. The continuum level is at $D$ = 0, the zero
level (no transmitted light) at $D$ = 1. One remarkable feature is the
rapid decrease of the mean absorption with decreasing redshift. In the
hierarchical models this is mostly due to the expansion of the universe
which decreases the gas density and the neutral fraction.  Excellent
agreement with the observed distribution is attained for both models,
after scaling the optical depth globally. It is worth pondering that
this is the result of a one-parameter fit which is equivalent to
adjusting the area under the curve giving the cumulative distribution.
The agreement between the shapes of the distribution is an independent
bit of information, telling us that these models do indeed produce a
Ly$\alpha$ forest with a realistic intensity distribution.  The
Eulerian $\Lambda$CDM model is doing slightly better than the SPH SCDM
simulation, in that the latter has more absorption at low column
densities, leading to a steeper slope than observed. This may reflect
differences in the way the heating during reionization was
incorporated.

Having obtained the scalefactors for the optical depths in the
simulations we can derive a lower limit on $\Omega_b$. For an
independent estimate of the radiation background intensity we adopt a
temporally constant UV background with  $\Gamma > 7\times10^{-13}$
s$^{-1}$.  This value corresponds to $J_{\nu} \approx 2.3 \times
10^{-22}$ (depending on the spectral shape). It is a {\it lower limit}
in that it includes only the radiation background from known QSOs alone.
From the firm
lower limit on $\Gamma$ due to QSOs alone we can determine a lower
limit for the amount of baryons in the universe:

\centerline{$\Omega_{baryon}h^{2}$ $>$ 0.017}

Our measurement (described in more detail in \cite{rau}) supports a high $\Omega_{baryon}$ (low D/H) universe,
still consistent with the upper range permitted by the solar system
light element abundances \cite{hata} and with the $\Omega_{baryon}$ derived from 
the D/H measurements by Tytler et al. (these proceedings).
Remaining uncertainties stem from the assumed intensity of the ionizing
radiation background, and the temperature the lower column density gas
has attained by the time reionization is complete.  Moreover, the
data sample is still relatively small, especially at lower redshifts.

The validity of these results depends of course not only the correct
cosmological model, but also on its technical realisation: how can we be sure
that we are not missing structure on very small scales, which are not
resolved by the simulations ? If there were clumpiness on sub-kpc
scales (unresolved by current techniques), the gas could
be locally more neutral and a smaller amount of baryons could
conceivably account for the same amount of opacity in the Lyman
$\alpha$ forest. From some point on such structure will diverge from
the properties of the CDM based cosmological models \cite{wein}
but the only way to find out for sure whether it exists is
to search directly for clumpiness in the gas responsible for most of
the Ly$\alpha$ forest absorption.

\section{Density Gradients on Sub-Kpc Scales}

The presence or absence of density and velocity structure on very small scales 
at high z can best be tested for by 
searching for differences between the absorption systems in adjacent
lines of sight to multiple (lensed) QSO images. We have embarked on a project 
to (re-)observe high redshift lensed QSOs with the Keck HIRES spectrograph, among them
the two cases UM 673 and HE 1104-1805, studied in the important papers
by Smette et al. (\cite{sme1,sme2}. The difference to earlier work is that with Keck
we are able to resolve the line widths even of most metal absorption systems, 
and can measure the column densities directly. 

\begin{figure}[htp]
\centerline{
\psfig{file=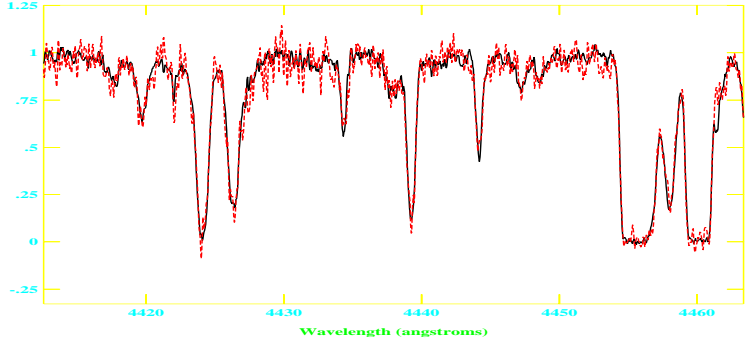,width=12.cm,height=5.cm,angle=-90.}
}
\caption{\small Ly$\alpha$ forest near z=2.64 in the line of sight to UM673 (transverse
separation 0.11 $h_{50}^{-1}$ kpc). The solid line is the spectrum of the A image,
the dotted one that of the B image. The only discernible difference between the
spectra comes from an MgI line at z=0.563  on the RHS of the rightmost 
Ly$\alpha$ complex only visible in the A image (separation 13.9 $h_{50}^{-1}$ kpc)\label{hispec}}
\end{figure}

The column density differences over a known separation in the plane of the sky
can be translated into gradients in the baryon density. We have completely 
profile-fitted a Lyman $\alpha$ forest region around z= 2.6 in both
the $A$ and $B$ images of the QSO UM673 and measured the column differences 
among those Voigt profile components agreeing in velocity position to within 10 kms$^{-1}$.
We restricted the analysis to 
HI column densities between $10^{12}$ and $10^{14}$ cm$^{-2}$, a range contributing
most prominently to the opacity at redshifts 2.6. This way we should be able to answer
our question whether there is structure in the gas producing the Ly$\alpha$
forest which may be capable of upsetting the $\Omega_b$ results.
It has been argued \cite{cro,hui} that the HI column density N(HI) should depend on the baryon density 
as $$ N(HI)\propto \rho^{\alpha}$$ with $\alpha \approx$ 1.6. For the RMS scatter of the logarithmic baryon density gradient we then obtain 
\begin{eqnarray}
\left<\left(\frac{d\log \rho}{dr}\right)^2\right> \leq \alpha^{-2} \left<\left(\frac{|\log N_A - \log N_B|}{dr}\right)^2\right>
\end{eqnarray}
The inequality arises because the variable thickness of the clouds
introduces an additional scatter in the column densities, even for a
density field constant everywhere.  Retaining only those line
pairs with a measurement error of less than 0.2 in the logarithmic column
density  differences  we arrive at an upper limit to the RMS estimate for fluctuations in the logarithmic
overdensity, \begin{eqnarray} \sqrt{\left<\left(\Delta\log
\rho\right)^2\right>} \leq 3.6\times10^{-2},\ \ 12.0 \leq {\rm N(HI)} \leq 14.0\ {\rm cm}^{-2}\end{eqnarray} valid
for a sample with a mean transverse separation of 0.16
h$_{50}^{-1}$ kpc between the lines of sight (for $q_0$=0.5).  Other choices of sample size give similar values.
Obviously, those parts of the universe contributing most to the 
Ly$\alpha$ forest opacity do not show detectable clumpiness, at least on the
scale we have now searched. Future investigations of lenses with somewhat
wider image separation and/or wavelength coverage further to the blue
will eventually provide information on the kpc range. 

\section{Velocity and Density Fields in Protogalaxies}

\begin{figure}[tb]
\centerline{
\psfig{file=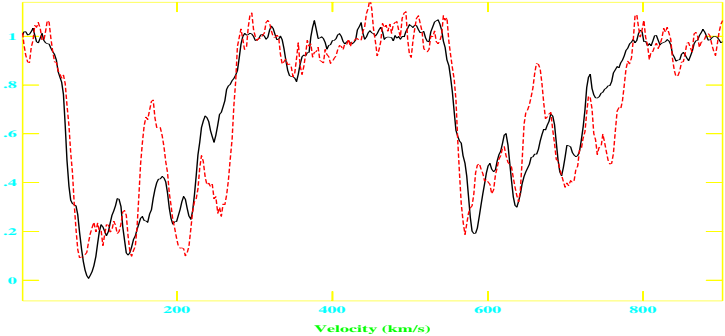,width=12.cm,height=5.cm,angle=-90.}
}
\caption{\small CIV ($\lambda\lambda$ 1548,1551 \AA ) complex at z=2.35 towards UM673 (LOS
separation 0.68 $h_{50}^{-1}$ kpc). \label{civspec}}
\end{figure}

Moving to higher column densities the lines of sight ultimately must be sampling
the gaseous extent of high redshift galaxies. The presence of these can be  
inferred from the striking variations between metal absorption lines in the different
images, which  stand out quite dramatically from the quiescent lower column density 
Ly$\alpha$ forest. Column densities even in the high ionization CIV gas can fluctuate
by a factor two over a few hundred parsecs (Fig. \ref{civspec}). Apparently, we
are seeing variations in the ionization parameter of the early interstellar medium. These variations can be used
to establish the sizes (more correctly the coherence lengths) for the gas units in those
galaxies, and the internal velocity dispersion.

\begin{figure}[tb]
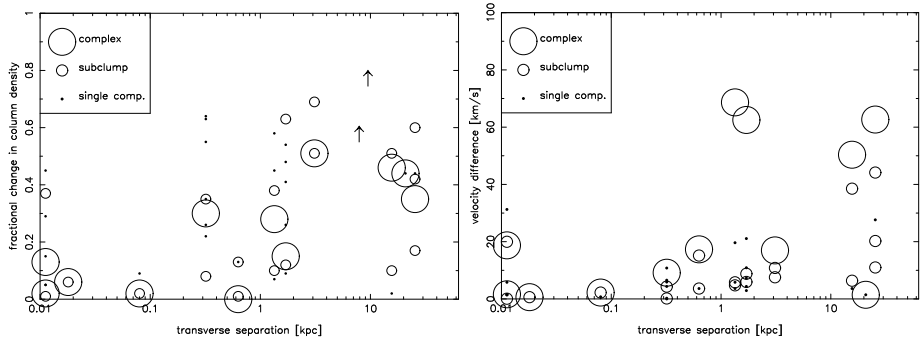

\centerline{
\psfig{file=rauchF7.ps,width=6.cm,angle=-90.}
\psfig{file=rauchF8.ps,width=6.cm,angle=-90.}  
}
\caption{\small Fractional difference in column density  (left), and the difference
in the column density weighted velocity (right) between the
two lines of sight as a function of transverse separation. Large circles represent whole CIV complexes (arrows are lower limits), small circles 
"subclumps", and the dots stand for single components. 
\label{civvnstat1}}
\end{figure}

Fig. \ref{civvnstat1} shows for the first time, how the decoherence in
the properties of these CIV absorption systems takes place in individual
lines of sight as a function of beam separation. 
In this figure each CIV complex (i.e., the full absorption system) is shown
by a large circle. "Subclumps", i.e., groups of lines, between which the continuum
recovers, have been assigned by eye somewhat subjectively, and are denoted
by small circles. Individual absorption components, as far as their fate can be
traced across the lines of sight, are represented by dots. The different classes
of symbols are of course not statistically independent as the single components
are parts of the subclumps etc.
It is obvious from the LHS figure that variations in the column density of individual
CIV components
have reached 50\%  at separations of a few hundred pc, while the total
complexes differ at that level once separations between a few and 20 kpc are reached. 
Velocity differences increase  to about 60 km$^{-1}$ at
separations of a few kpc (RHS plot). As one goes to the very smallest separations the differences both in
velocity and column density do not diminish as much as expected for a naive 
extrapolation to zero separation at the QSO redshift itself. This may be indicating
that the associated ($z_{abs} \approx z_{em}$) clouds are smaller than intervening
systems,  do not cover
the QSO continuum emission region, or possess some internal motion on very small
scales. 

We can use the
differences in column density as a function of separation to estimate,
in a statistical sense, the size of the CIV absorbing galaxies as a
function of the column density detection threshold.  Making the very
crude assumption that all these objects have identical sizes and
densities, and approximating their column density -- radius relation by
a Gaussian, we find a mean radius of 14 h$_{50}^{-1}$ kpc at the
3.7$\times$10$^{12}$cm$^{-2}$ ($\sim$ 15 m\AA\ equivalent width) contour. We emphasize that this
is a very crude estimate as it appears already from the few systems observed
that there is a whole range of sizes. Interestingly, this coherence length 
for high ionization gas (at z$\approx$ 2--3) is
considerable smaller than than that for MgII (low ionization) systems at low
redshifts and similar detection threshold ($\sim$ 180 h$_{50}^{-1}$ kpc for 0.5 $<$ z
$<$ 1.3; \cite{womb}). The increase in cross-section could be due to
galaxy halos growing, or an increasing metal enrichment  of the universe by volume,
as time proceeds.

%

%


%

%


\acknowledgements{ 
The $\Omega_b$ measurement is the result of an ongoing collaboration
with J. Miralda-Escud\'e, W.L.W.  Sargent, T. A. Barlow, D.H.,
Weinberg, L. Hernquist, N. Katz, R. Cen, and J.P. Ostriker, to be
published in greater detail in the ApJ Nov.  1997 issue.  The work on
lenses is done in collaboration with  W.L.W. Sargent and T. A.
Barlow. The observations were made at the W.M. Keck Observatory which
is operated as a scientific partnership between the California
Institute of Technology and the University of California; it was made
possible by the generous support of the W.M.  Keck Foundation.  I am
grateful to M. Steinmetz for providing the column density plot from his
simulation (fig. 2), and to NASA for support through grant
HF-01075.01-94A from the Space Telescope Science Institute.
}


%
\vfill
\end{document}